\documentclass[prl,aps,epsf,showpacs,twocolumn]{revtex4}
\usepackage{times}
\usepackage{graphicx}
\usepackage{float}
\usepackage{latexsym,amsmath,amssymb,bm,euscript}
\usepackage{color}
\usepackage{subfigure}
\usepackage{epstopdf}
\usepackage[colorlinks=true,linkcolor=blue,citecolor=blue]{hyperref}
\usepackage{hyperref}

\begin{document}

\title{Competing Spin Liquid States in the Spin-$1/2$ Heisenberg Model On Triangular Lattice}

\author{Wen-Jun Hu} 
\author{Shou-Shu Gong}
\email{shoushu.gong@gmail.com}
\author{Wei Zhu}
\author{D. N. Sheng}
\affiliation{Department of Physics and Astronomy, California State University, Northridge, California 91330, USA}


\begin{abstract}

We study the spin-$1/2$ Heisenberg model on the triangular lattice with the antiferromagnetic first ($J_1$)
and second ($J_2$) nearest-neighbor interactions using density matrix renormalization
group. By studying the spin correlation function, we find a $120^{\circ}$ magnetic order phase for $J_2 \lesssim 0.07 J_1$
and a stripe antiferromagnetic phase for $J_2 \gtrsim 0.15 J_1$. Between these two phases, we identify a spin liquid region characterized by the exponential decaying spin and dimer correlations, as well as 
the large spin singlet and triplet excitation gaps on finite-size systems. We find two near degenerating ground states with distinct properties in two sectors, which indicates more than one spin liquid candidates in this region. While the sector with spinon is found to respect the time reversal symmetry,
the even sector without a spinon breaks such a symmetry for finite-size systems.
Furthermore, we detect the signature of the fractionalization
by following the evolution of different ground states with 
inserting spin flux into the cylinder system. Moreover,
by tuning the anisotropic bond coupling, we explore the nature of the spin liquid phase and find the optimal
parameter region for the gapped $Z_2$ spin liquid.

\end{abstract}


\maketitle

Quantum spin liquids (SLs) are long-range entangled states
with remarkable properties of fundamental importance \cite{Balents2010}.
The SL physics has been considered to be essential to
understand strongly correlated systems and unconventional
superconductivity \cite{Anderson1973, Lee2006}. The 
simplest and perhaps most striking SLs are the gapped
topological SLs, which develop a topological order \cite{Wen1989, Wen1990, wen1990topological}
with the emergent fractionalized quasiparticles \cite{Wen1991, Senthil2000, Senthil2001}.
Although SLs have been studied intensively for two decades
and demonstrated in contrived models \cite{Rokhsar1988, Read1991,
Moessner2001, Balents2002, Senthil2002, Lesik2002, Sheng2005, Kitaev2006,
Yao2007, Greiter2007, Melko2011}, the microscopic condition for the emergence 
of SLs in frustrated magnetic systems is not well understood.

At the experimental side, possible SLs have been discovered in
various materials. Among these materials,
the most promising systems are the kagome antiferromangets
including the Herbertsmithite and Kapellasite \cite{mendels2007, helton2007, vries2009, fak2012, han2012},
as well as the organic Mott insulators with a 
triangular lattice structure such as $\kappa$-(ET)$_2$Cu$_2$(CN)$_3$ \cite{shimizu2003, kurosaki2005, yamashita2008, yamashita2009}
and EtMe$_3$Sb[Pd(dmit)$_2$]$_2$ \cite{itou2008, yamashita2010}.
In all these materials, no magnetic order is observed at the temperature much
lower than the interaction energy scale.
These experimental findings have inspired intensive theoretical studies on the frustrated
magnetic systems with strong frustration or competing interactions.

Theoretically, the kagome Heisenberg model appears to possess
a robust SL. Density matrix renormalization group (DMRG)
studies suggest a gapped SL \cite{jiang2008, yan2011, depenbrock2012, jiang2012nature},
which may be  consistent with a $Z_2$ topological order \cite{depenbrock2012, jiang2012nature}.
Variational studies based on the projected fermionic parton
wave functions however favor a gapless Dirac SL \cite{ran2007, iqbal2013, iqbal2014}. 
Interestingly, by introducing the second and third neighbor
couplings \cite{messio2012, gong2014kagome, he2014csl} or the chiral interactions \cite{bauer2014},
DMRG \cite{gong2014kagome, he2014csl, bauer2014} studies
recently discovered another topological SL --- chiral
spin liquid (CSL) \cite{kalmeyer1987, wen1989csl}, which breaks time
reversal symmetry (TRS) spontaneously and is identified as the $\nu = 1/2$
bosonic fractional quantum Hall state. On the other hand, the non-magnetic phases
in the frustrated honeycomb and square $J_1$-$J_2$ models appear to be
conventional valence-bond solid state \cite{Ganesh2013, zhuzhenyue2013, Gong2013, Gong2014square}.

The spin-$1/2$ triangular nearest-neighbor antiferromagnetic (AF) Heisenberg model was the first
candidate proposed to realize a SL
ground state by Anderson \cite{Anderson1973}, although it turns out to still exhibit
a $120^\circ$ AF order \cite{sachdev1992, bernu1992, sorella1999, zheng2006, white2007}.
To understand the triangular weak Mott insulator materials,
combined theoretical and numerical studies \cite{lesik2005, sheng2009, matthew2011}  on a spin model
with four-site ring-exchange couplings \cite{Misguich1999}
find a gapless \textit{spin bose metal} with spinon Fermi sea.
To enhance frustration \cite{bacci1990, chubukov1992, manuel1999, mishmash2013, kaneko2014, campbell2015,
yunoki2006, sheng2006, leon2010, white2011, li2014},
one way is to include the second-neighbor
coupling $J_2$, where a stripe ordered state is found with larger $J_2$ coupling \cite{bacci1990, chubukov1992}, and an intermediate non-magnetic region may emerge \cite{manuel1999, mishmash2013, kaneko2014, campbell2015} based on different approaches.
The variational Monte Carlo simulations find a nodal d-wave
SL \cite{mishmash2013} and a gapless SL \cite{kaneko2014}
as the candidates for this intermediate phase, although such method has the enhanced tendency of finding
fractionalized phase in nonmagnetic region. Very recently, a DMRG work \cite{zhu2015} found the
indication of a gapped SL which conserves the TRS in the non-magnetic phase. However, the nature of the
quantum phase with the intermediate $J_2$  remains far from clear.

\begin{figure}[t]
 \includegraphics[width=1.0\linewidth]{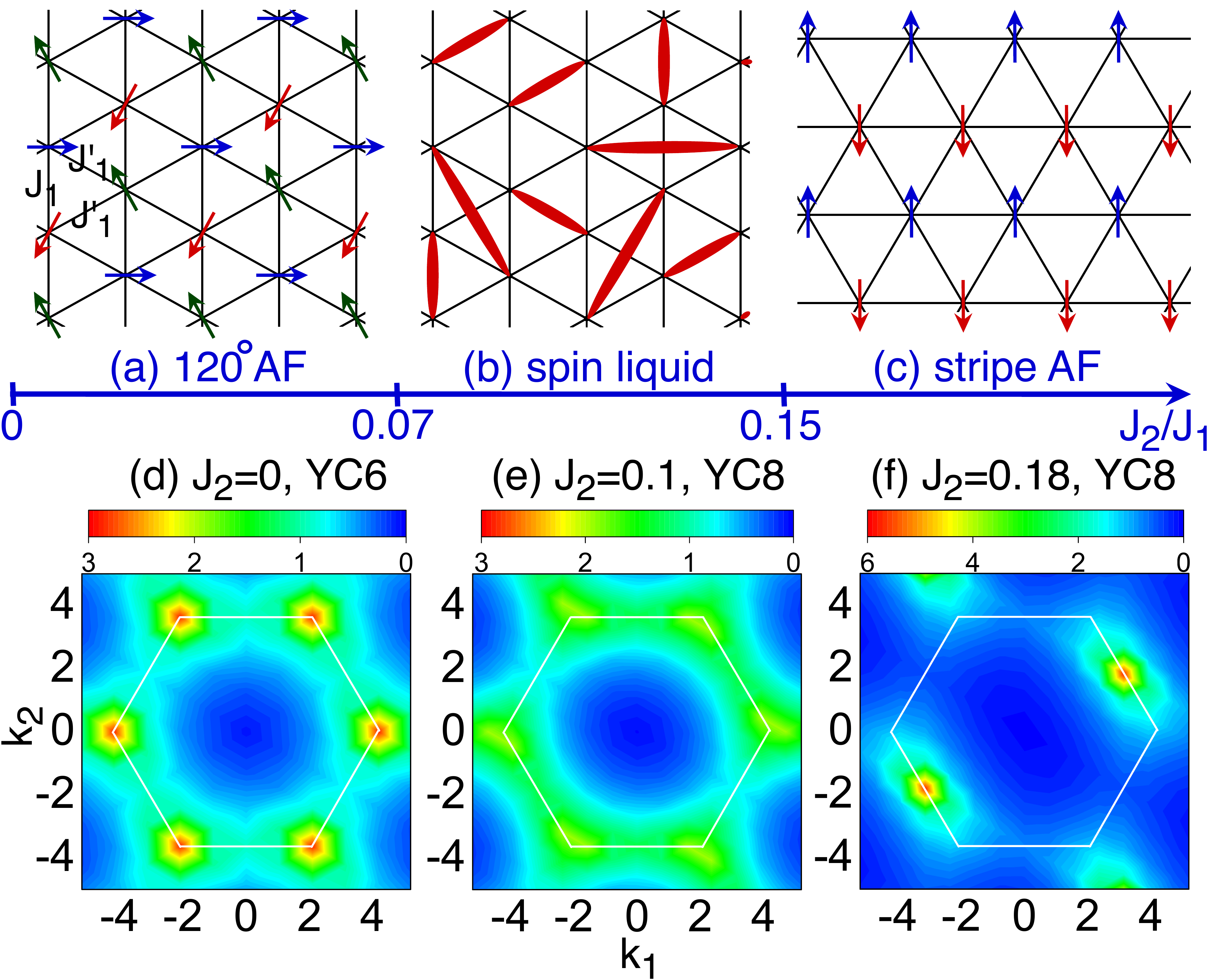}
 \caption{(color online) Quantum phase diagram of the isotropic spin-$1/2$ $J_1$-$J_2$ Heisenberg model on triangular
	  lattice ($J_1=J'_1$). With growing $J_2$, the system has a $120^{\circ}$ AF phase for $J_2 \lesssim 0.07$, a
	  stripe AF phase for $J_2 \gtrsim 0.15$, and a  SL phase in between.
	  The schematic figures of the different phases also show the YC (a,b) and XC (c) cylinder geometries.
	  (d)-(f) are the contour plots of spin structure factor for each phase.}\label{phase}
\end{figure}

In this Letter, we study the spin-$1/2$ triangular model with
the AF first and second nearest-neighbor $J_1(J'_1)$-$J_2$ couplings based on DMRG calculations.
The model Hamiltonian is given as
\begin{equation}
 H = J_1 \sum_{\langle i,j \rangle_{\rm vertical}} \vec{S}_i \cdot \vec{S}_j 
 + J'_1 \sum_{\langle i,j \rangle_{\rm zigzag}} \vec{S}_i \cdot \vec{S}_j 
  + J_2 \sum_{\langle\langle i,j \rangle\rangle} \vec{S}_i \cdot \vec{S}_j,\nonumber
\end{equation}
where the sums $\langle i,j \rangle$ and $\langle\langle i,j \rangle\rangle$
run over all the first- and second-neighbor bonds,
respectively. The first-neighbor couplings $J_1$ and $J'_1$ are for the vertical 
and zigzag bonds as shown in Fig.~\ref{phase}(a). We study most systems with $J'_1 = J_1$
unless we specify otherwise.
We set $J_1 = 1$ as the energy scale. By studying the spin correlations,
we find a non-magnetic region sandwiched by a $120^\circ$ AF phase with three sublattices
for $J_2 \lesssim 0.07$ and a stripe AF phase for $J_2 \gtrsim 0.15$ as shown in Fig.~\ref{phase}.
In this non-magnetic region, we identify two ground states
with distinct properties in two sectors, indicating two competing candidates for SL phases.
The spin and dimer correlations decay exponentially with small correlation lengths.
Interestingly, the chiral correlations decay exponentially fast for the ground state
in the odd sector with an edge spinon, while it develops the long-range correlations in the even sector 
(with no spinon) for finite-size systems, consistent with the level crossing between two SLs for
the systems with different boundaries.
The fractionalized spinon is detected through adiabatically inserting spin flux.
While the state in the odd sector agrees with a TRS preserving SL,
the TRS breaking SL (e.g., chiral SL) may be a competing or nearby state 
in more extended parameter space.
Moreover, the strong anisotropy of bond energy along different directions
is observed for some finite-size systems, which may imply a nematic order
for gapped $Z_2$ SL \cite{xu2014}.
This possible $Z_2$ SL is observed to be stabilized by a small bond coupling anisotropy ($J'_1 \gtrsim  J_1$), which suppresses chiral order in both sectors.

We study the cylinder systems using 
highly  accurate $SU(2)$ DMRG \cite{white1992, mcculloch2002} for most of calculations and 
$U(1)$ DMRG \cite{white1992}
for inserting flux \cite{gong2014kagome}.
Two cylinder geometries known as  XC and YC are studied, which have one lattice
direction parallel to the $x$ or  $y$ axis as shown in Fig.~\ref{phase}.
We denote them as XC$L_y$-$L_x$ (YC$L_y$-$L_x$), where $L_y$ and $L_x$
are the number of sites along the $y$ and $x$ directions, respectively. 
We study the cylinder systems with $L_y$ up to $10$ lattice spacings by keeping
up to $20000$ $U(1)$-equivalent states in $SU(2)$ DMRG and $5000$ states
for inserting flux. The truncation errors are less than $10^{-5}$ in all
calculations, which leads to accurate results.

\textit{Even and odd topological sectors.---}
Based on the resonating valence-bond picture, the ground states of SL
on cylinder can be either in the even or odd sector
according to the parity of the number of bonds that are
cut by a vertical line along the enclosed direction.
Usually, the odd sector can be obtained by removing or adding
one site on each open edge of cylinder, which has been used successfully
to find different sectors of the gapped SLs in kagome systems \cite{yan2011, he2014,
zhu2014}. 

\begin{figure}[t]
 \includegraphics[width=0.8\linewidth]{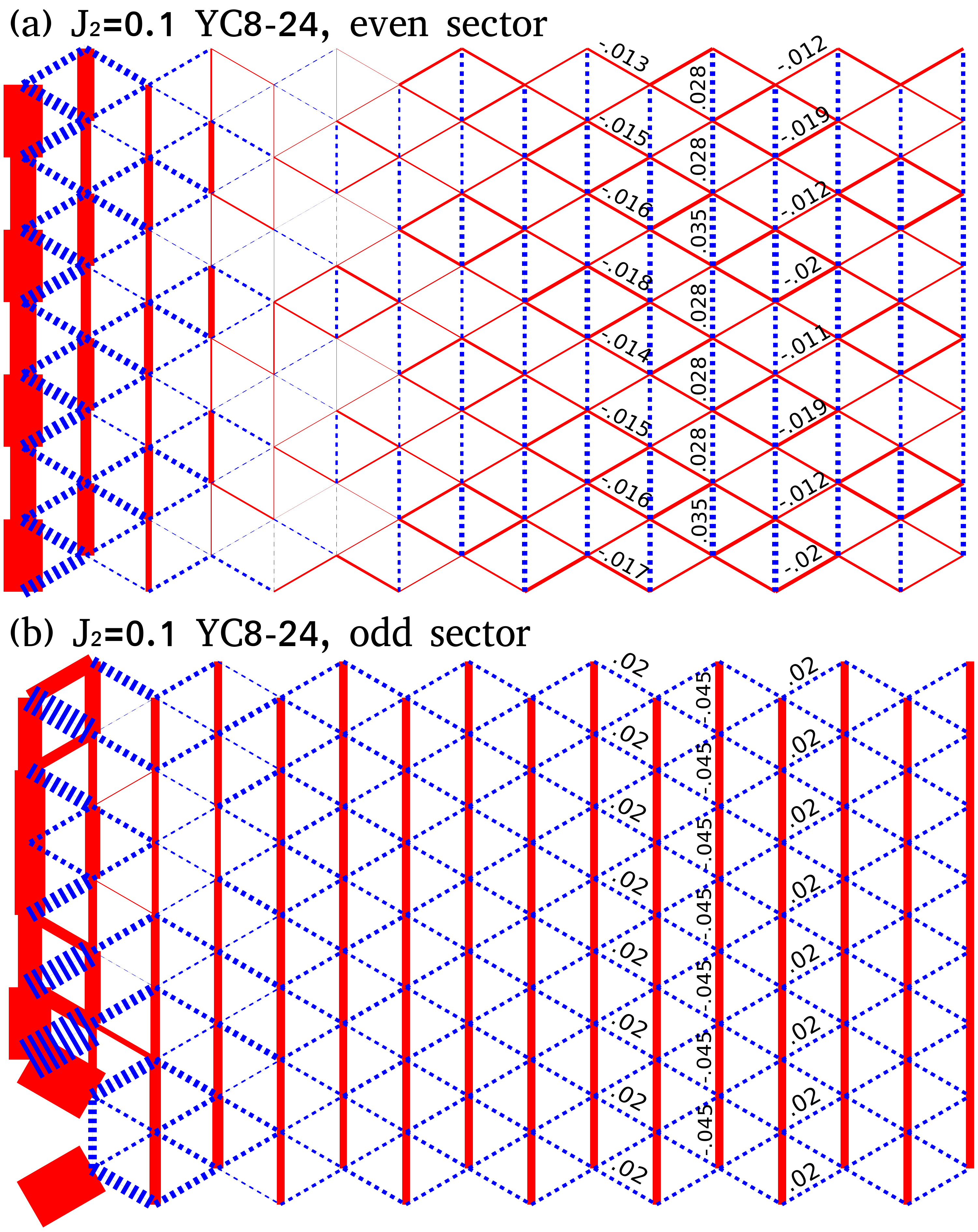}
 \caption{(color online) The NN bond energy  $\langle S_i \cdot S_j\rangle$ for $J_2 = 0.1$ on the YC8-24 cylinder in 
          (a) the even and (b) the odd sector. The left $16$ columns are shown here. The odd sector is obtained by removing one site in
	  each boundary of cylinder. In both figures, all the bond energy have subtracted the average value $-0.18$. 
	  The red solid and blue dashed bonds denote the negative and positive bond energies after subtraction 
	  (with some numbers shown for clarity).}\label{sector}
\end{figure}

By doing simulations without and with removing one site at each boundary,
we always find different ground states in these two sectors on YC cylinders ($L_y=6,8,10$),
which are shown in Fig.~\ref{sector} for YC8 cylinder as an example (See 
Supplemental Material for YC10 cylinder) \cite{suppl}.
We find that the two sectors have the different bond energy distributions
in the bulk of cylinder. While the vertical bonds
are weaker in the even sector, they become the stronger ones
in the odd sector as shown in Fig.~\ref{sector}. 
The nematic order, which is defined as the difference
between the strong and weak bonds to describe the lattice rotational
symmetry breaking, exhibits the distinct behaviors for the two states
on our studied systems. While the nematic order grows with increased cylinder width in the odd
sector, it decreases in the even sector \cite{suppl}.

\begin{figure}[t]
 \includegraphics[width=1.0\linewidth]{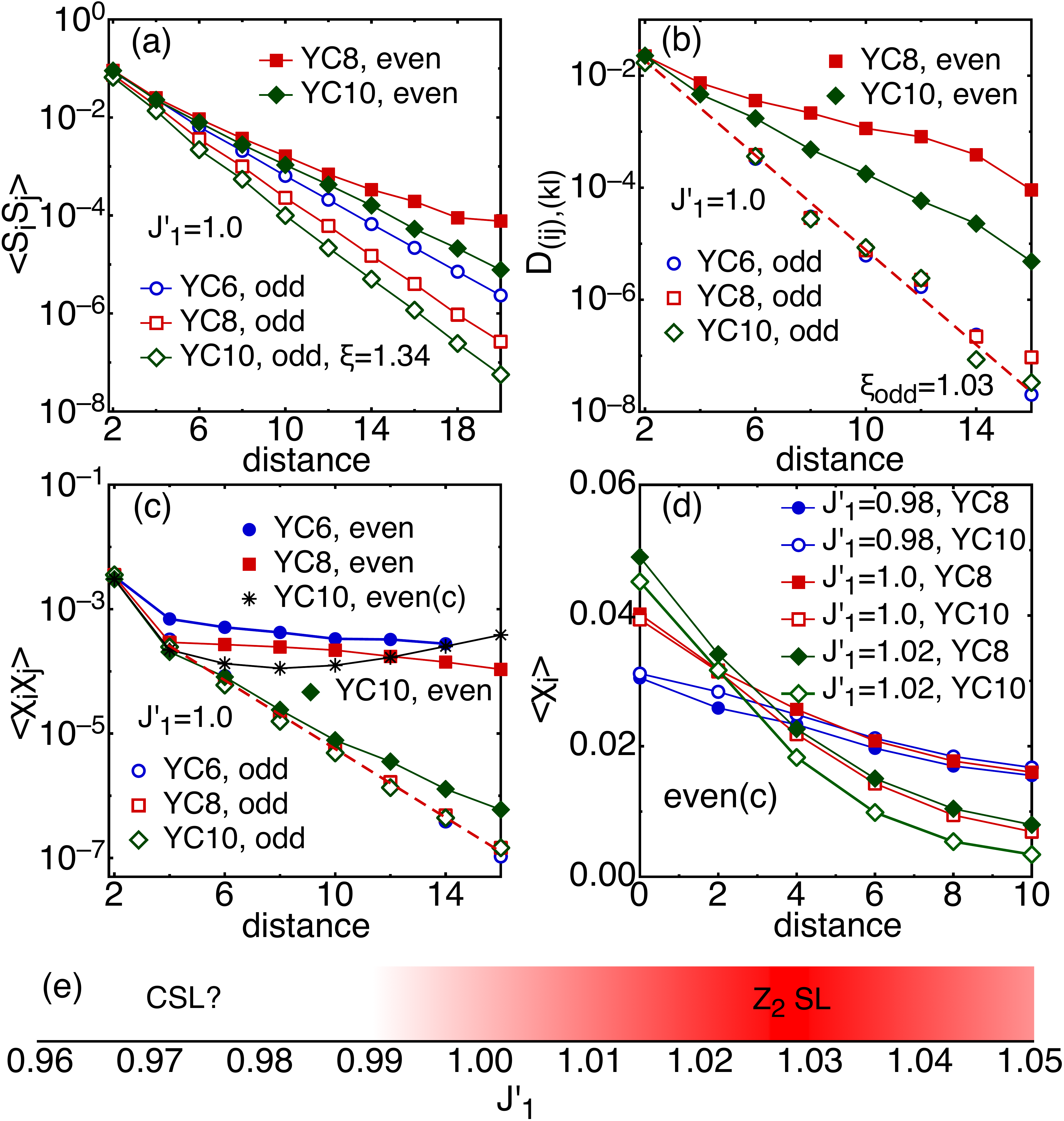}
 \caption{(color online) (a) and (b) are the spin and dimer correlation functions on different YC cylinders for $J_2 = 0.1$.
	  $\xi$ denotes the corresponding correlation lengths on YC10 cylinder in the odd sector.
	  (c) Chiral correlations for $J_2 = 0.1$ on different cylinders. All the data are obtained from real number DMRG 
	  calculations except ``YC10, even(c)'', which denotes the data obtained from complex wave function
	  DMRG calculations on YC10 cylinder in the even sector. (d) Chiral order from the boundary to the bulk for 
	  the bond anisotropic system in the even sector, which are obtained from the complex DMRG. (e) Schematic phase diagram
	  for the bond anisotropic system at $J_2=0.1$.}\label{measure}
\end{figure}

\textit{Characteristic properties of different SL states.---}
Next, we further characterize the two states by studying
correlation functions. In Fig.~\ref{measure}(a)
we show the spin correlations for $J_2 = 0.1$, which decay
faster with growing cylinder width in both states, 
indicating the vanishing spin order in both states.
In Fig.~\ref{measure}(b), we demonstrate the dimer correlation
function $D_{(ij),(kl)} = \langle (S_i \cdot S_j)(S_k \cdot S_l) \rangle
-\langle S_i \cdot S_j \rangle \langle S_k \cdot S_l \rangle$,
which also decay exponentially to vanish.  Interestingly,  the states
in the odd sector have  very short correlation lengths 
almost independent of the system width.  But in the even sector,
the correlation lengths are longer than those in the odd
sector for the smaller $L_y=6$ and $8$, which decrease with growing system width.

To study the possible TRS breaking, we calculate the
scalar chiral correlation function $\langle \chi_{i} \chi_{j}\rangle$
($\chi_{i} = (S_{i,1} \times S_{i,2}) \cdot S_{i,3}$). In both sectors,
the chiral order has the same pattern,
where the up and down triangles have the same chirality direction.
As shown in Fig.~\ref{measure}(c), in the odd sector, the
chiral correlations decay quite fast to vanish without developing
long-range order. However, the chiral order and spontaneous TRS breaking are very robust for
the $L_y=6,8$ systems in the even sector at the intermediate phase, where additional two
fold ground state degeneracy is also obtained in the DMRG simulation for each system
(with opposite chirality) associated with the TRS breaking.
As we increase system width to $L_y=10$, the chiral correlation becomes
less robust, where different results are obtained depending on if we use
complex or real initial wave function in DMRG simulation.
The chiral correlation remains to be long-ranged in the complex wave function,
where the TRS is spontaneously broken. However, if we use
real number wave function, the DMRG will
find a state with short-range chiral correlations (the real state has near identical
bulk energy as the complex state but higher energy near the edge).

To further clarify the chiral order in the system, we consider a bond 
anisotropy perturbation by tuning the nearest-neighbor zigzag bond strength
as $J_{1}^{\prime}$ (see Fig.~\ref{phase}(a)). For $J_2 = 0.1$,
we find that the SL region  persists for $0.95 < J_{1}^{\prime} < 1.05$.
In the odd sector, the chiral order vanishes, and all the properties are consistent
with $J_{1}^{\prime} = 1.0$. In the even sector, the chiral order appears stronger for $0.95 < J_{1}^{\prime} \lesssim 0.99$,
where it grows a bit from YC8 to YC10 cylinder (see Fig.~\ref{measure}(d)
for $J_{1}^{\prime} = 0.98$). For $J_{1}^{\prime} \gtrsim 1.0$, the chiral order decays pretty fast
from the boundary to the bulk especially for the larger YC10 cylinder,
which indicates  a possible vanishing of the chiral order in this region in the thermodynamic limit.
At $J_1^{\prime} = 1.0$, the chiral correlations are strong and show long-range behavior, but the chiral order also decays
with the increase of system width.
Thus, the two states in both sectors may recover TRS at large system limit
for $1.0 \lesssim J_{1}^{\prime} < 1.05$, which is the most possible region for
stabilizing a $Z_2$ SL. On the other hand, for $0.95 < J_{1}^{\prime} \lesssim 0.99$, chiral order
becomes stable in the even sector, while the fate of such a phase remains unclear depending on
if the chiral order would develop in the spinon sector in the thermodynamic limit.
We illustrate our  finding in the phase diagram Fig.~\ref{measure}(e).

\begin{figure}[t]
 \includegraphics[width=1.0\linewidth]{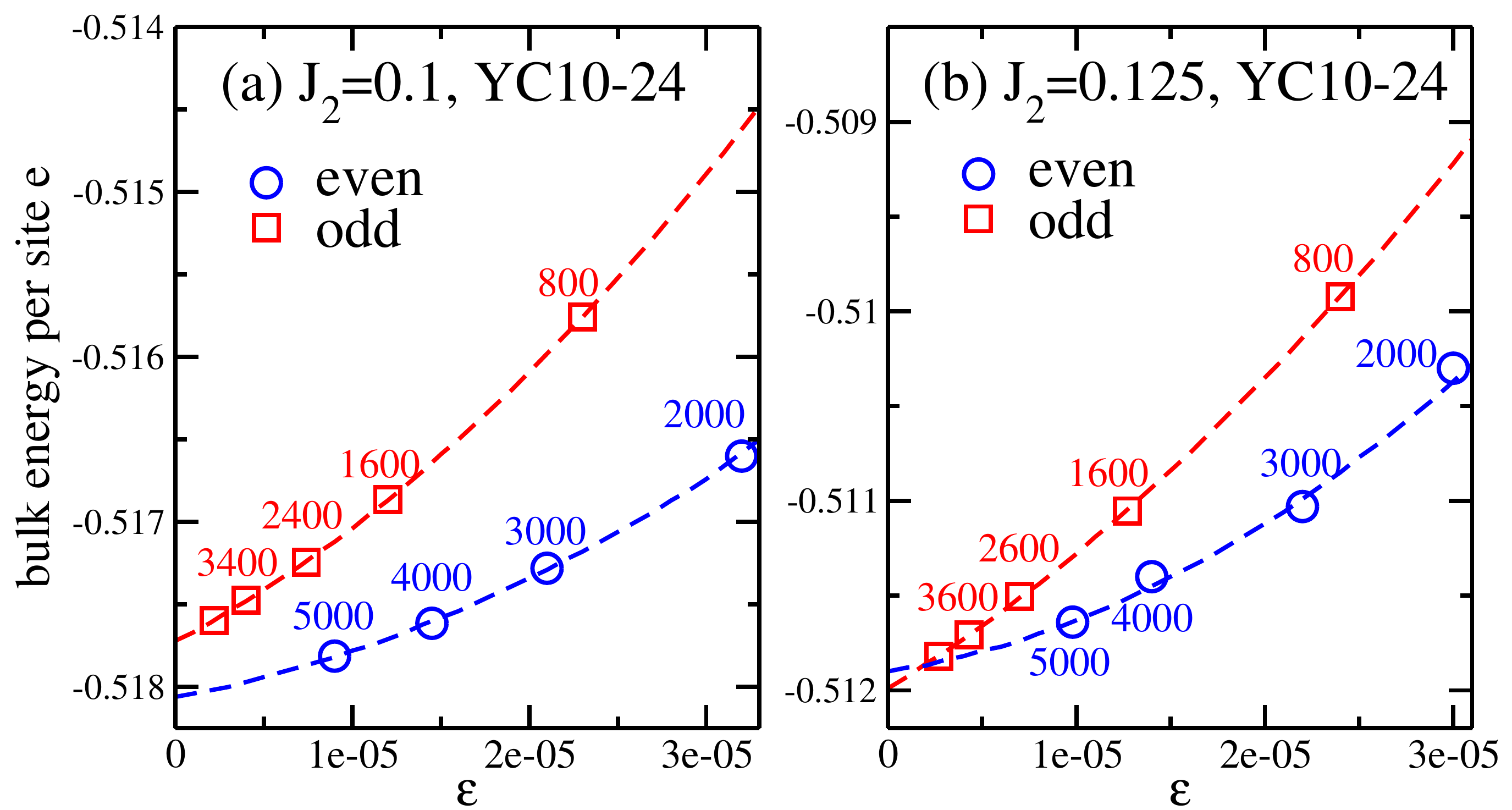}
 \caption{(color online) Bulk energy per site $e$ versus DMRG truncation error $\varepsilon$ for the ground states
	  in the even and odd sectors for (a) $J_2 = 0.1$ and (b) $J_2 = 0.125$ on the YC10-24 cylinder.
	  The numbers denote the kept $SU(2)$ states. The energy data are fitted using the formula 
	  $e(\varepsilon)=e(0) + a\varepsilon + b\varepsilon^2$.}\label{scaling}
\end{figure}

We calculate the bulk ground-state energy in both sectors.
The energy per site for different systems are presented 
in  Table~\ref{table} for $J_2 = 0.1, 0.125$.
For the smaller system widths ($L_y=6$ and $8$), the odd sectors generally have
the lower energy than the  even sectors,
which lead to a positive  energy splitting
$\Delta e= e_{\rm even} - e_{\rm odd}$. 
However, this splitting drops very fast with  the increase of  $L_y$, which is tiny
for system width $L_y=10$ (for example $\Delta e \simeq -0.0004$ for $J_2=0.1$,
see Fig.~\ref{scaling}) indicating the close energy for states in both sectors.
We also compute the singlet $\Delta_s$ and triplet $\Delta_T$ gaps
by obtaining the ground state first and then sweeping the two low-lying states simultaneously
or the $S=1$ sector in the bulk of cylinder \cite{white2011}. The results are shown in Table~\ref{table}.

\begin{table}
\begin{tabular}{|c|c|c|c|c|c|}
\hline
$J_2$,YC$L_y$ & $e_{\rm even}$ & $e_{\rm odd}$ & $\Delta e$ & $\Delta_{\rm T}$ & $\Delta_{\rm S}$ \\
\hline
$0.1$,YC6 & $-0.5155$ & $-0.5210$ & $0.0055$ & $0.365$ & $0.30$ \\	
\hline
$0.1$,YC8 & $-0.5171$ & $-0.5195$ & $0.0024$ & $0.335$ & $0.26$ \\
\hline
$0.1$,YC10 & $-0.5181(2)$ & $-0.5177$ & $-0.0004(2)$ & $0.30(1)$ & $0.18$ \\
\hline
$0.125$,YC6 & $-0.5104$ & $-0.5145$ & $0.0041$ & $0.389$ & $0.33$ \\
\hline
$0.125$,YC8 & $-0.5115$ & $-0.5133$ & $0.0018$ & $0.343$ & $0.22$ \\
\hline
$0.125$,YC10 & $-0.5119(2)$ & $-0.5120$ & $0.0001(2)$ & $0.30(1)$ & $0.15$ \\
\hline
\end{tabular}
\caption{The bulk energy per site in the even ($e_{\rm even}$) and odd ($e_{\rm odd}$) sectors, the energy difference
	 $\Delta e = e_{\rm even} - e_{\rm odd}$, the spin triplet ($\Delta_{\rm T}$) and
	 singlet ($\Delta_{\rm S}$) gaps  in the odd sector for $J_2 = 0.1$ and $0.125$ on the YC$L_y$ ($L_y=6,8,10$)
	 cylinders. We use the fully converged results for $L_y = 6$ and $8$, and the extrapolated results for $L_y=10$ as 
	 shown in Fig.~\ref{scaling}.}\label{table}
\end{table}

\begin{figure}[t]
 \includegraphics[width=0.7\linewidth]{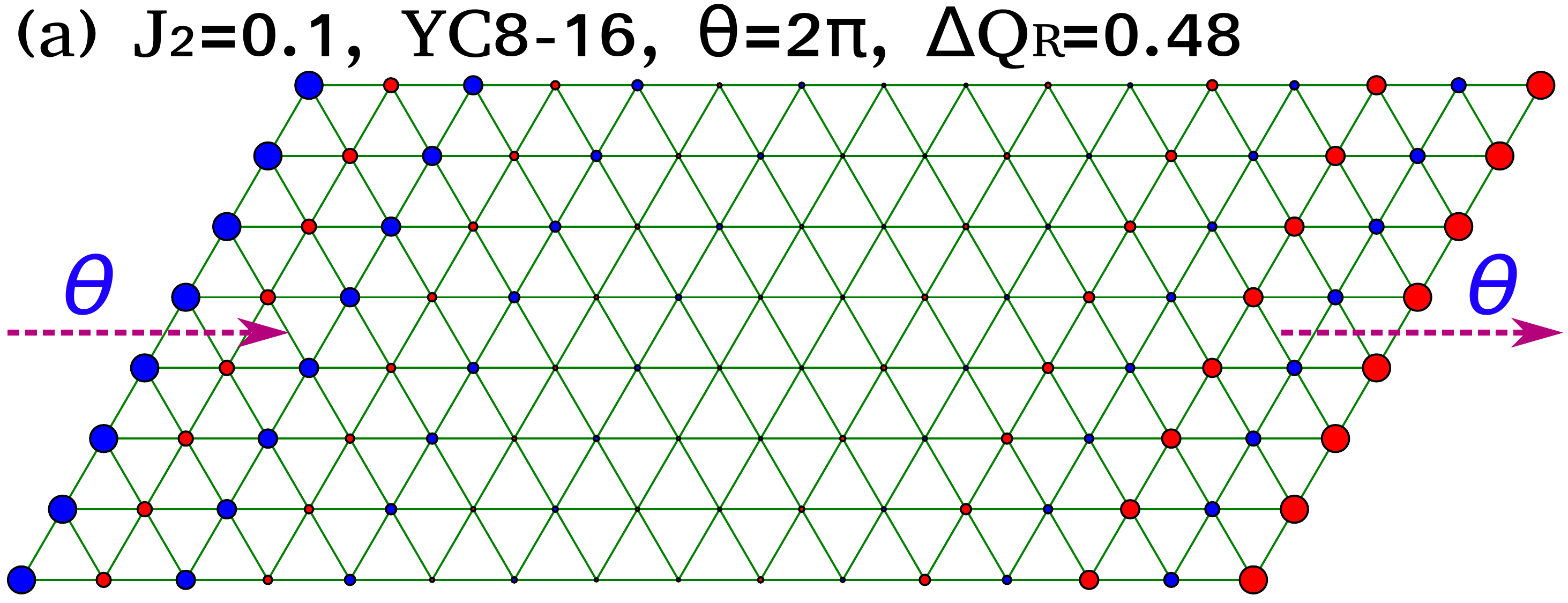}
 \includegraphics[width=0.7\linewidth]{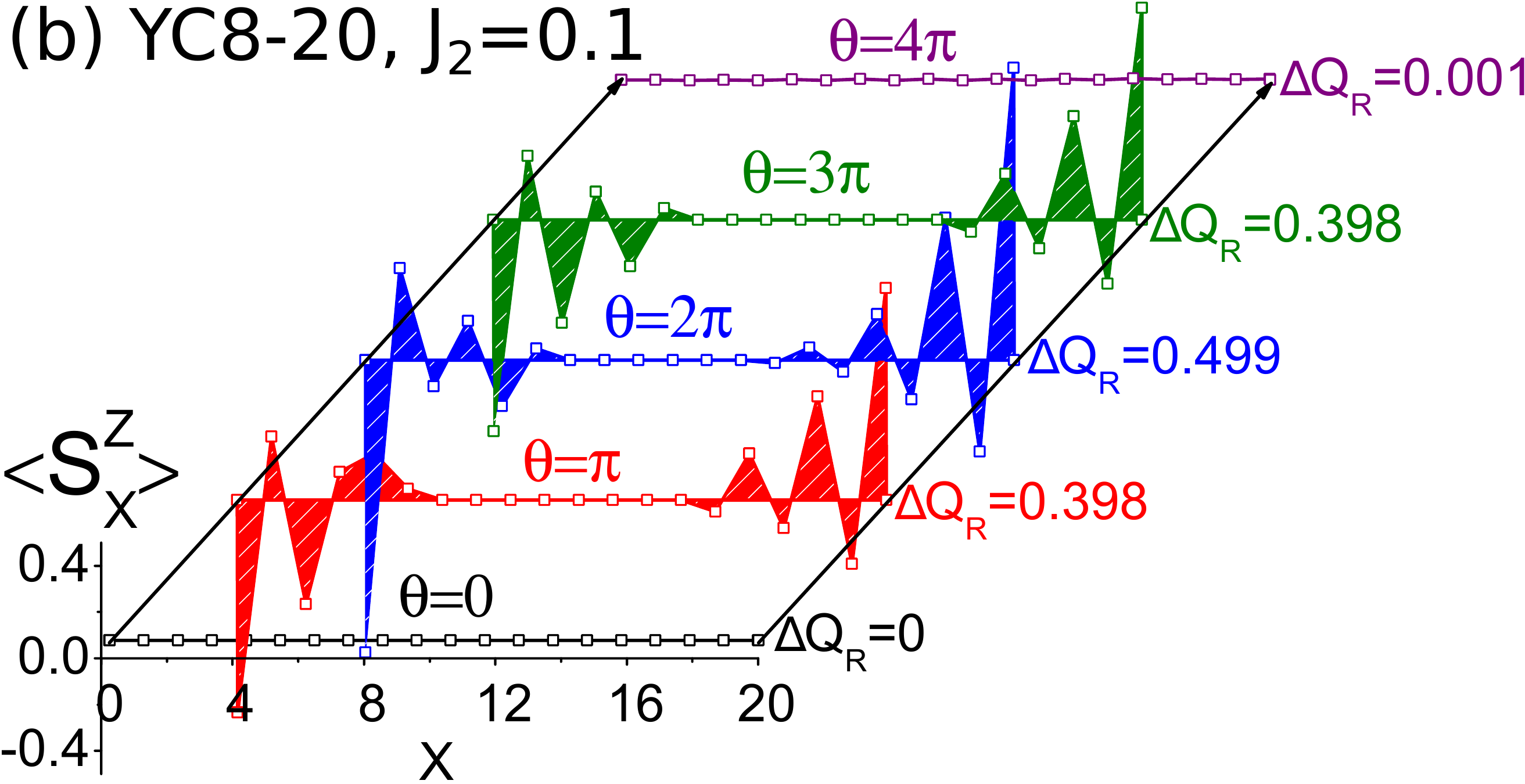}
 \includegraphics[width=0.6\linewidth]{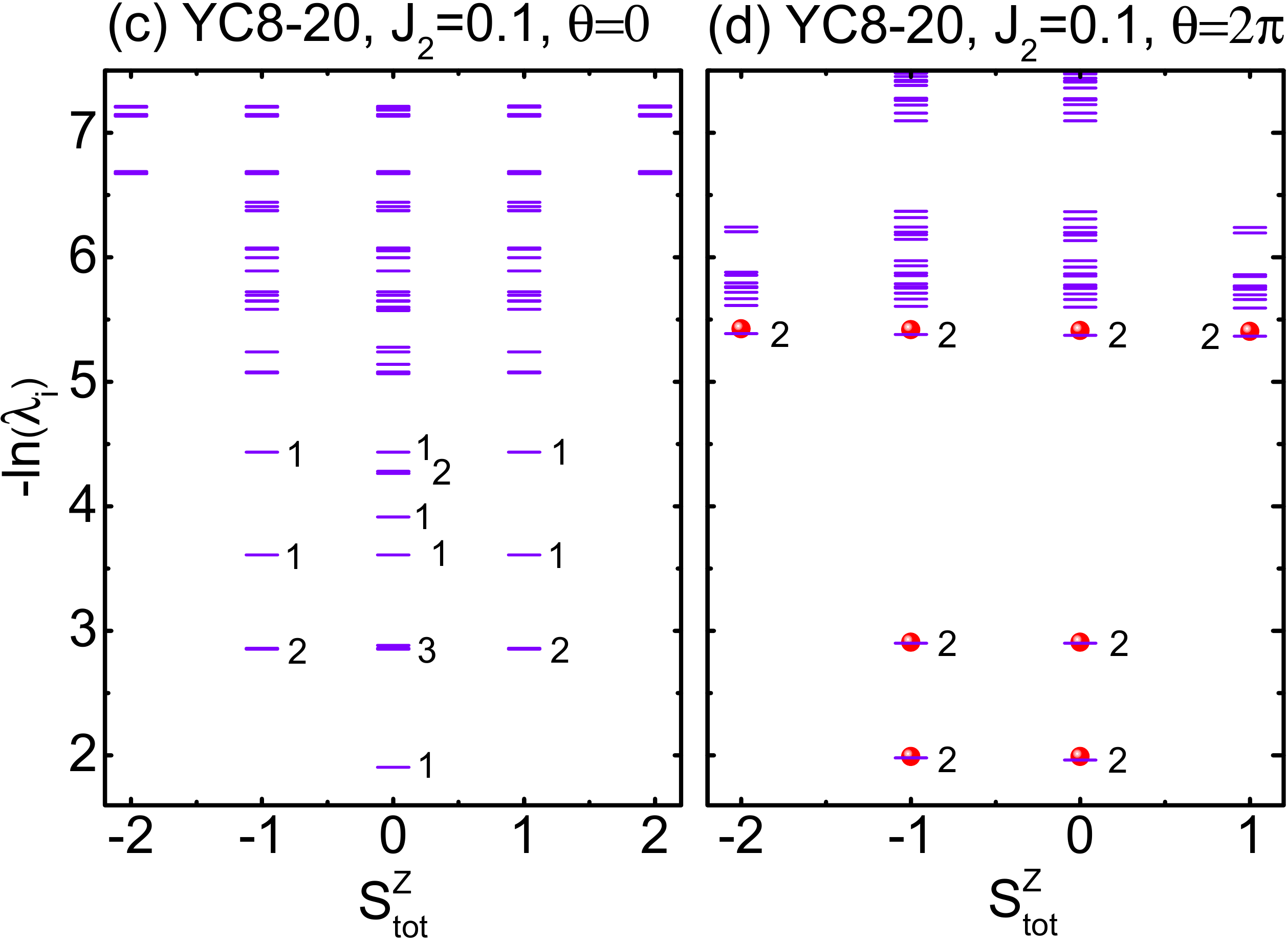}
 \caption{(color online) (a) Real-space configuration of spin magnetization
	  $\langle S^{z}_{i} \rangle$ after adiabatically inserting a
	  flux $\theta = 2 \pi$. The red and blue circles denote the positive and
	  negative $\langle S^{z}_{i} \rangle$ with the area of circle proportional
	  to the amplitude of $\langle S^{z}_{i} \rangle$. $\Delta Q_{\rm R}$ is 
	  the net spin accumulation on the right edge $\Delta Q_{\rm R} = \sum_{i} \langle S^{z}_i \rangle$,
	  where $i$ denotes the site on the right edge of the cylinder.
	  (b) Real-space configuration
	  of the accumulated spin magnetization in each column with increasing flux.
	  Low-lying ES at the flux (c) $\theta = 0$ and (d) $\theta = 2 \pi$. The numbers
	  in ES denote the near-degenerate eigenvalues in the large weight levels. At $\theta = 2 \pi$
	  in (d), all the eigenvalues are double degenerate as explicitly shown by 
          the circles for the low-lying levels.}\label{flux}
\end{figure}

\textit{Inserting flux and  the  nature of different sectors.---}
Inserting flux is an effective way to find different
sectors and to determine the basic properties of the
quasi-particles in the ground states,
which has been applied in DMRG to study different 
topological SLs \cite{he2014, gong2014kagome, zhu2014}.
To introduce a flux, we impose the twist boundary condition
in the $y$ direction by replacing terms 
$S^{+}_i S^{-}_j + h.c. \rightarrow S^{+}_i S^{-}_je^{i\theta} + h.c.$ 
for all neighboring bonds crossing the $y$-boundary.
We start from the even sector by adiabatically
increasing  $\theta$ and measuring the evolution of
the spin-$z$ local magnetization $\langle S^z_i \rangle$.
With increasing $\theta$, a net spin-$z$ accumulates on the open edges
as shown in Fig.~\ref{flux}(a), which indicates that
the quasiparticle responding to the flux here must carry spin,
such as the spinon in chiral spin liquid \cite{gong2014kagome} and the fermionic spinon (spinon bonded with
vison) in $Z_2$ SL \cite{he2014, qi2015, zaletel2015}. 

With the flux $\theta = 0 \rightarrow 2 \pi$,
the net spin grows continuously from $0$ to $0.5$ on the edges
as shown in Fig.~\ref{flux}(b).
At $\theta = 2 \pi$, an $S^z=\pm 1/2$ spinon develops on each boundary,
and the ground state evolves to a new sector.
By  further increasing  $\theta = 2 \pi \rightarrow 4 \pi$,
the net spin dissipates continuously to zero, and the system
evolves back to the even sector.  
In Figs.~\ref{flux}(c) and \ref{flux}(d), we demonstrate the entanglement spectrum (ES) with
inserting flux. At $\theta = 0$ in the even sector, the whole
ES is symmetric about $S^z = 0$. At $\theta = 2 \pi$, the ES
evolves to be symmetric about $S^z = 1/2$, which is consistent with
the observed fractionalized spin-carrying  quasiparticles on boundaries.
At $\theta = 4 \pi$, the ES becomes the same as that at
$\theta = 0$, indicating that the system evolves back to the even
sector. Interestingly, each eigenvalue of the ES at 
$\theta = 2 \pi$ is doubly degenerate. By comparing
the odd sector obtained by removing sites and the sector
with inserted flux $\theta = 2 \pi$, we find that these two states
have the same bulk energy and ES,
indicating that the two sectors obtained by different
methods are exactly the same.
This odd sector might be consistent with the
fermionic spinon sector of the $Z_2$ SL \cite{qi2015, zaletel2015, ashvin2015}.

\textit{Summary and Discussions.---}
By means of DMRG calculations on wide cylinder systems (up to $L_y=10$ lattice sites) of
the spin-$1/2$ $J_1$-$J_2$ Heisenberg model on the triangular
lattice, we find a SL region  bordered by a $120^{\circ}$ N\'{e}el AF phase at small $J_2 \lesssim 0.07$
and a stripe AF phase for $J_2 \gtrsim 0.15$. The spin and dimer correlations
are shown to decay fast for wider systems with small correlation lengths
comparable to lattice constant
with large spin and singlet excitation gaps.
The ES in the odd sector could be consistent
with the theoretical description of the fermionic spinon in the topological theory of $Z_2$ SL.
However, the long-range chiral correlation is observed to be strong in the even sector for a space isotropic model.
By tuning the anisotropic bond coupling, we find that the possible gapped $Z_2$ SL is stabilized by some weak
anisotropy ($J_1^{\prime} \sim 1.02 $). The chiral correlations are enhanced in the even sector by the opposite
anisotropy ($J_1^{\prime} \sim 0.98$), which may be stabilized in both sectors by TRS breaking terms and 
we leave this open issue for future studies.


We acknowledge the stimulating discussions with T.~Senthil,
X.~G.~Wen, Z.~Y.~Zhu, S.~R.~White, F.~Wang, and Y.~Qi. 
This research is supported by the National
Science Foundation through grants 
  PREM DMR-1205734 (W.J.H.)
and DMR-1408560 (S.S.G.), 
and the U.S. Department of Energy,
Office of Basic Energy Sciences under grant No. DE-FG02-06ER46305
(W.Z., D.N.S.).
\\

\textit{Note added.}---While completing this work, we became aware of
some related papers \cite{zhu2015, saadatmand2015}. We reached the similar
conclusion on gapped SL with Ref.~\cite{zhu2015}.

\bibliographystyle{apsrev}
\bibliography{main_text}{}

\clearpage

\section*{Supplemental Material: Competing Spin Liquid States in the Spin-$1/2$
Heisenberg Model On Triangular Lattice}

\subsection{I. Correlation functions}

\begin{figure}[t]
 \includegraphics[width=1.0\linewidth]{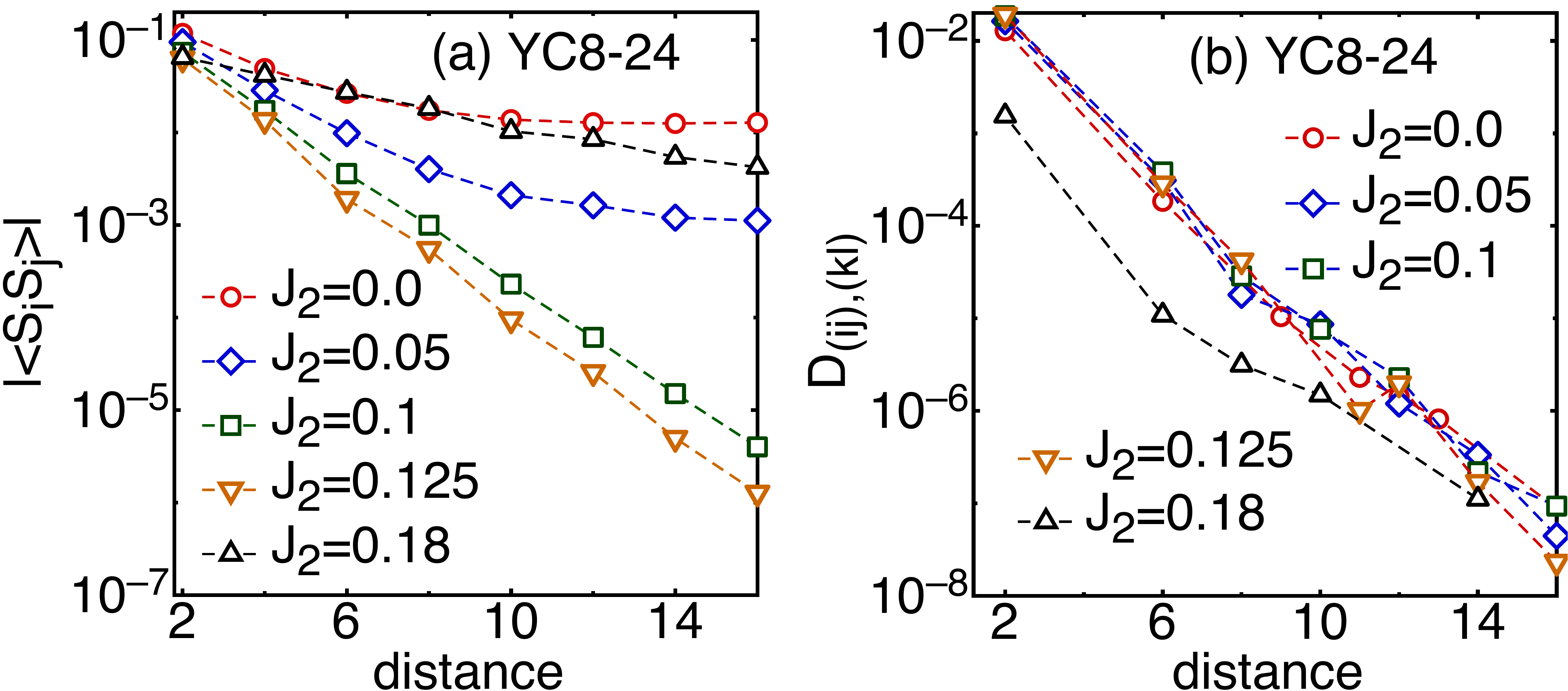}
 \caption{(color online) (a) and (b) are the spin and dimer correlation functions 
 for various $J_2$ couplings on the YC8-24 cylinder. In the spin liquid phase region,
 both correlation functions decay exponentially to vanish.
 }\label{suppl:spin}
\end{figure}

First of all, we demonstrate the spin and dimer correlation functions
for different $J_2$ on YC8-24 cylinder as shown in Fig. \ref{suppl:spin}.
In this figure, the correlation functions for $J_2=0.1$ 
and $0.125$ are shown in the odd sector.
In Fig.~\ref{suppl:spin}(a), the spin 
correlations $|\langle S_i \cdot S_j \rangle|$ decay slowly 
as a function of distance for $J_2=0$, $0.05$, 
and $0.18$, indicating the long-range magnetic order; instead for $J_2=0.1$ and $0.125$, 
it decays exponentially to vanish, which is consistent with the
absent magnetic order. 

\begin{figure}[b]
 \includegraphics[width=0.8\linewidth]{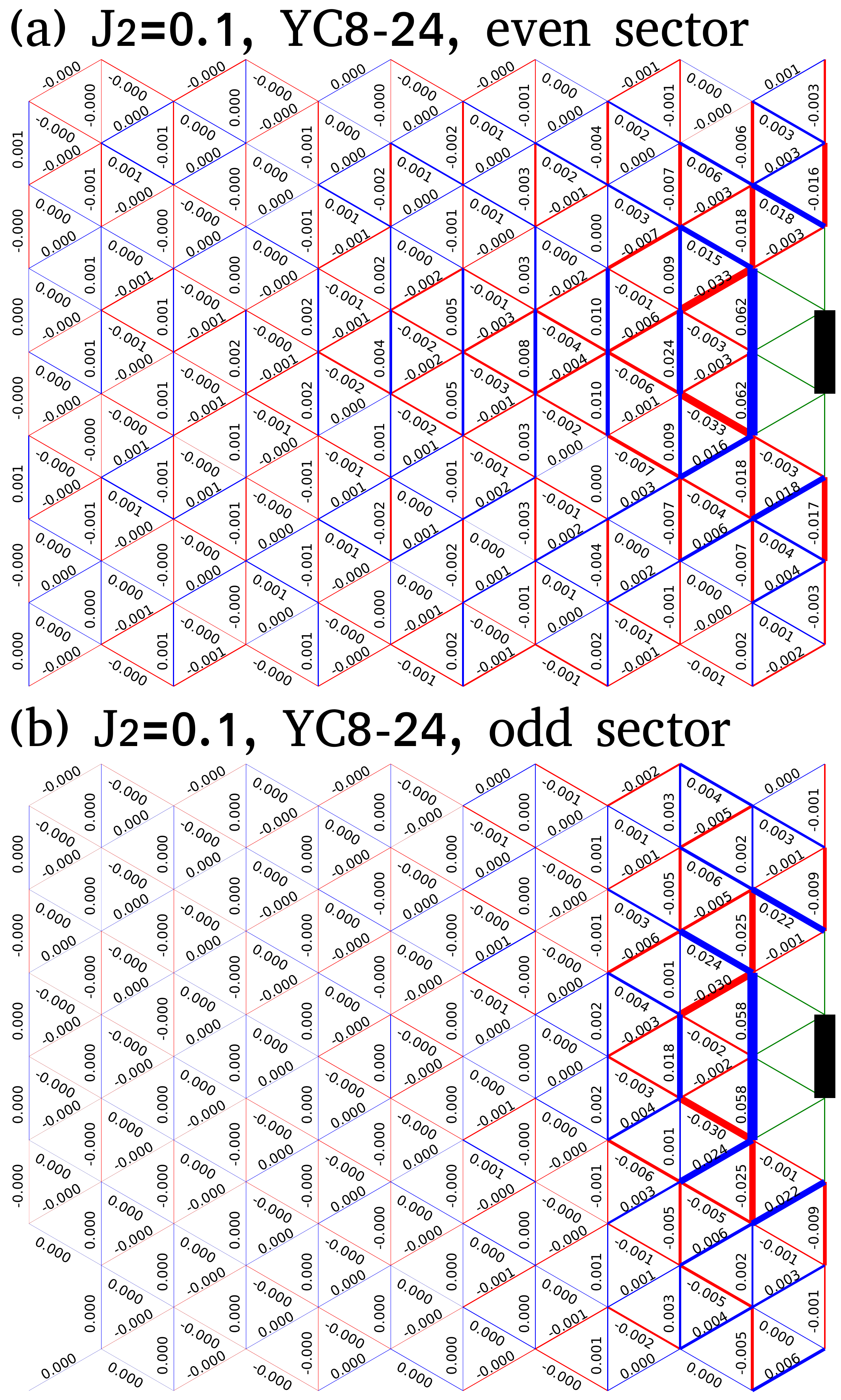}
 \caption{(color online) The dimer-dimer correlation function 
	  for $J_2 = 0.1$ on YC8-24 cylinder in (a) the even
	  sector and (b) the odd sector. The black bond in the middle is the reference bond. 
	  The blue and red bonds denote the positive and negative dimer correlations, respectively.}\label{suppl:dimer}
\end{figure}

In order to investigate the possible valence-bond solid order,
we study the dimer-dimer correlation function on cylinder systems,
which is defined as
\begin{equation}
 D_{(ij),(kl)} = \langle B_{ij}B_{kl} \rangle-\langle B_{ij}\rangle \langle B_{kl} \rangle,
\end{equation}
where $(i, j)$ and $(k, l)$ represent the nearest-neighbor (NN)
bonds, and $B_{ij}=S_i \cdot S_j$.
In Fig.~\ref{suppl:dimer} we show the dimer correlations
for $J_2=0.1$ on YC8-24 cylinder in both the even and odd sectors.
The dimer correlation in the odd sector decays faster
than that in the even sector, which is shown in Fig. 3(b) of the main text.
In Fig.~\ref{suppl:spin}(b), we show the dimer correlations for different $J_2$,
which all decay fast to vanish with an exponential manner.

\begin{figure}[t]
 \includegraphics[width=0.5\linewidth]{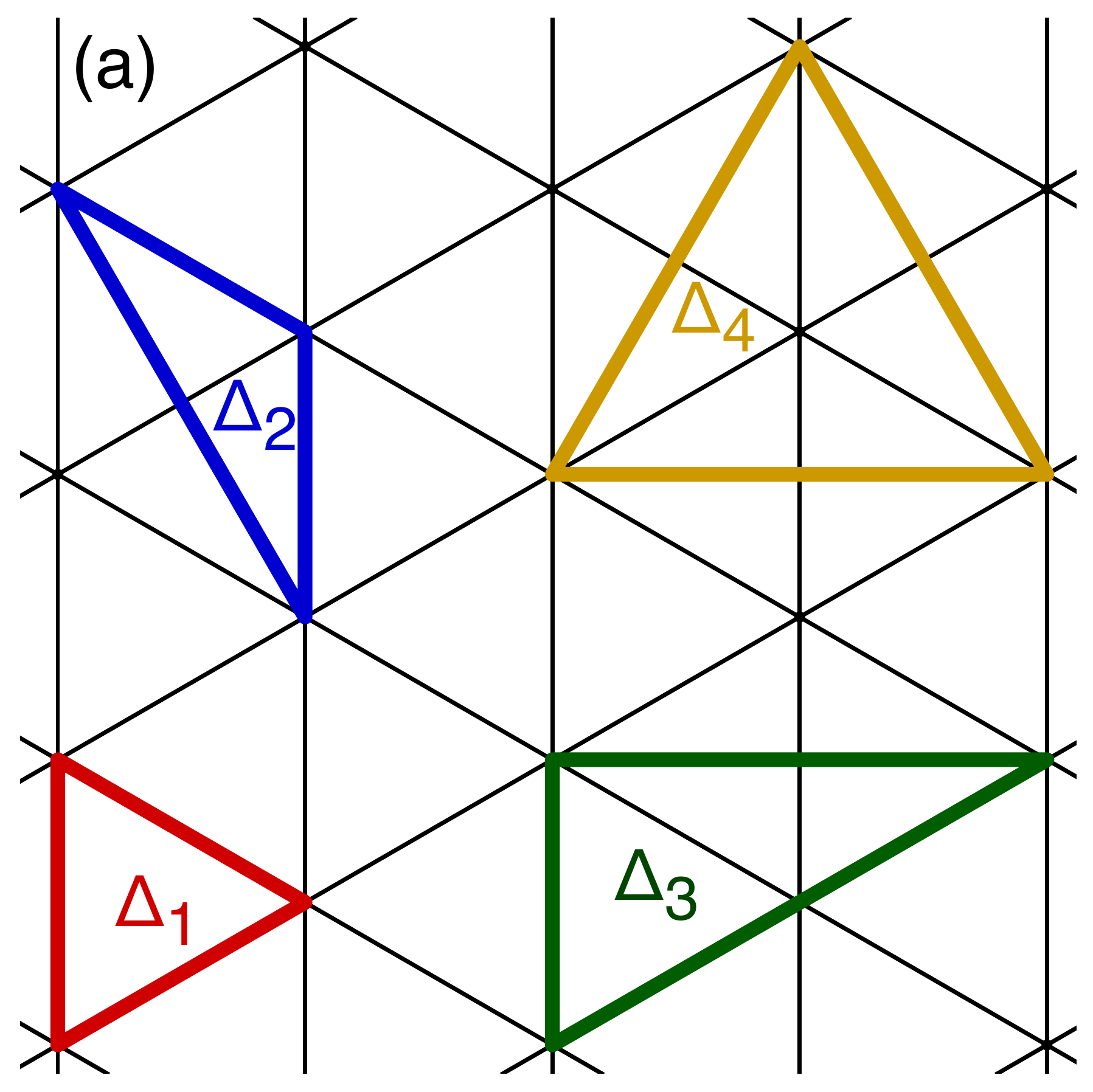}
 \includegraphics[width=1.0\linewidth]{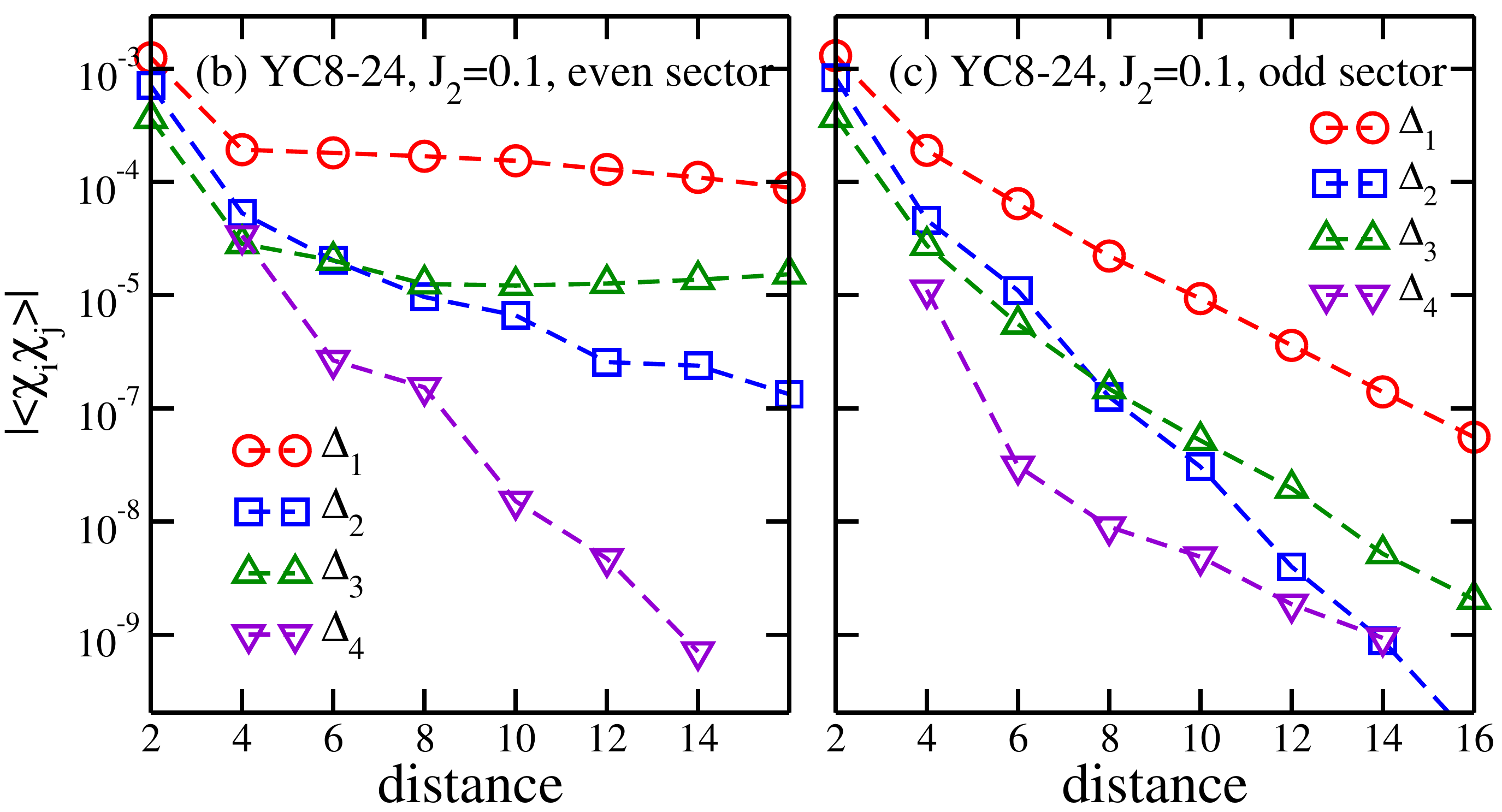}
 \includegraphics[width=1.0\linewidth]{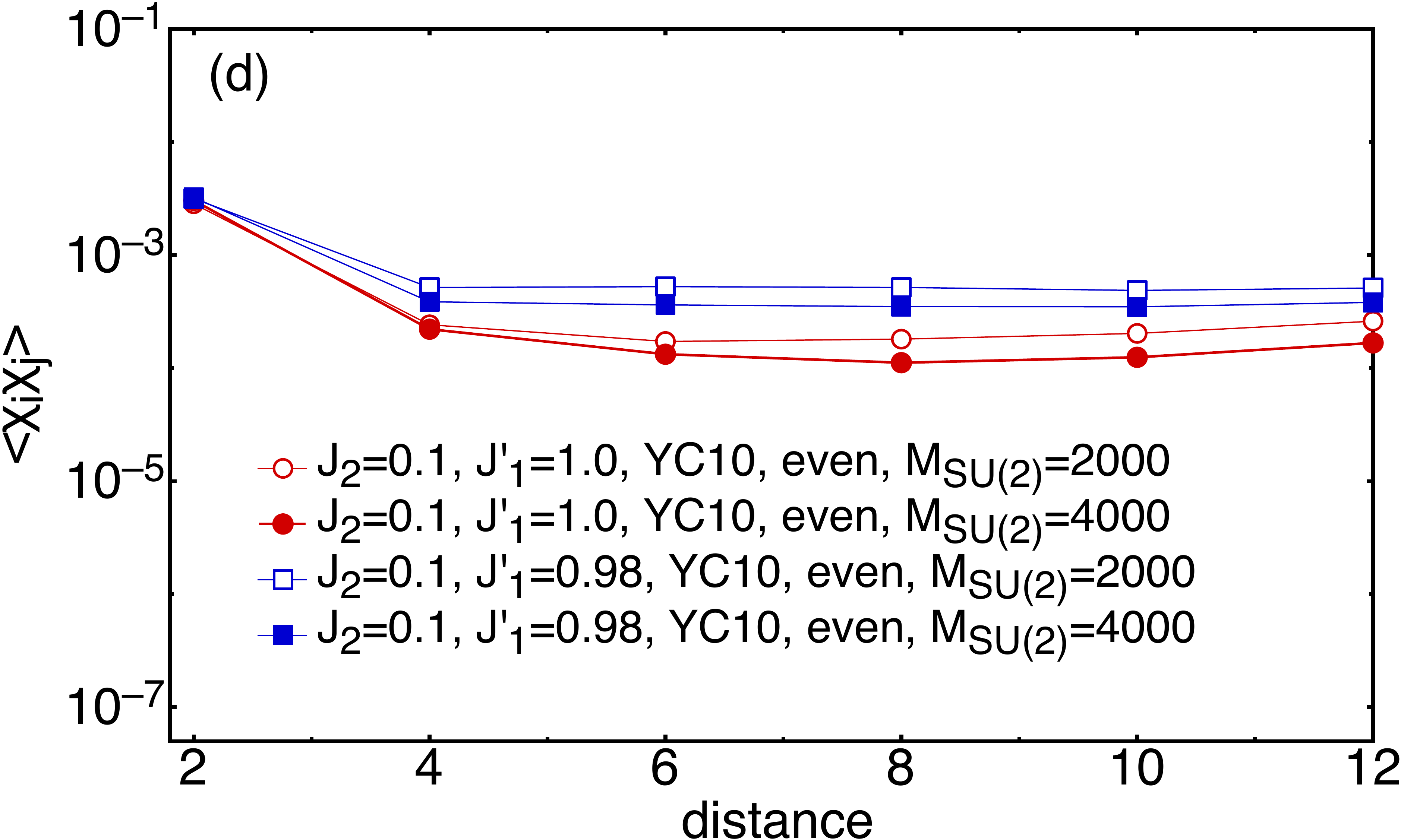}
 \caption{(color online) (a) The four three-spin triangles which are used 
          to calculate the chira-chiral correlation functions
	  in (b) and (c). (b) and (c) are the chiral-chiral correlation functions 
	  versus triangle distance for $J_2 = 0.1$ on YC8-24
	  cylinder in the even and odd sectors. These chiral correlations are obtained from
	  the real number DMRG. 
	  (d) is the chiral-chiral correlation functions of the triangles $\Delta_1$ at $J_2=0.1$ with anisotropic bond coupling $J'_1=1.0$ and $0.98$
	  in the even sector on YC10-20 cylinder with different kept $SU(2)$ DMRG states.}\label{suppl:chiral}
\end{figure}

To detect the possible time-reversal symmetry (TRS) breaking, we measure 
the chiral-chiral correlation functions for the four kinds of 
triangles $\Delta_1$, $\Delta_2$, $\Delta_3$, and $\Delta_4$ as shown
in Fig.~\ref{suppl:chiral}(a) in both the even and odd sectors.
The chiral-chiral correlation function is defined as
\begin{equation}
 \langle \chi_{i}\chi_{j}\rangle = \langle [S_{i,1}\cdot(S_{i,2}\times S_{i,3})]
 [S_{j,1}\cdot(S_{j,2}\times S_{j,3})] \rangle,
\end{equation}
where $\chi_{i}=S_{i,1}\cdot(S_{i,2}\times S_{i,3})$
is the scalar chiral order parameter of triangle $\Delta_{i}$.
As an example, we show the chiral correlations for $J_2 = 0.1$ on the YC8-24 cylinder
in Fig.~\ref{suppl:chiral}.
In the odd sector, all the chiral correlations decay quite fast to vanish (see Fig.~\ref{suppl:chiral}(c)),
which is also observed on YC6 and YC10 cylinders and indicates no TRS breaking.
However, in the even sector as shown in Fig.~\ref{suppl:chiral}(b), the chiral correlations
for the triangles $\Delta_1, \Delta_2$ and $\Delta_3$ exhibit the long-range chiral order, while those for
$\Delta_4$ decay fast. Using the complex number DMRG, we can
measure the chiral orders $\langle \chi_{i}\rangle$ for all the triangles directly.
In the odd sector, the chiral orders for all the triangles are vanished.
In the even sector, we find that the chiral orders for $\Delta_1, \Delta_2$ and $\Delta_3$ are finite,
and $\langle \chi_{i}\rangle$ for $\Delta_4$ is much smaller (for $J_2 = 0.1$ on YC10 cylinder, it is about $10^{-4}$).
In Fig.~\ref{suppl:chiral}(d) we show the chiral correlations of $\Delta_1$
by keeping different $SU(2)$ DMRG states for $J_2=0.1$,
$J'_1=1.0$ and $0.98$ on YC10 cylinder in the even sector.
By increasing the kept $SU(2)$ DMRG states from $2000$ to $4000$, the
chiral correlations change slightly, indicating the near convergence of the chiral correlations
with increasing states. The non-zero chiral orders indicate the TRS breaking in the even sector.

\subsection{II. Bond energy distributions in different sectors}

\begin{figure*}[b]
 \includegraphics[width=\linewidth]{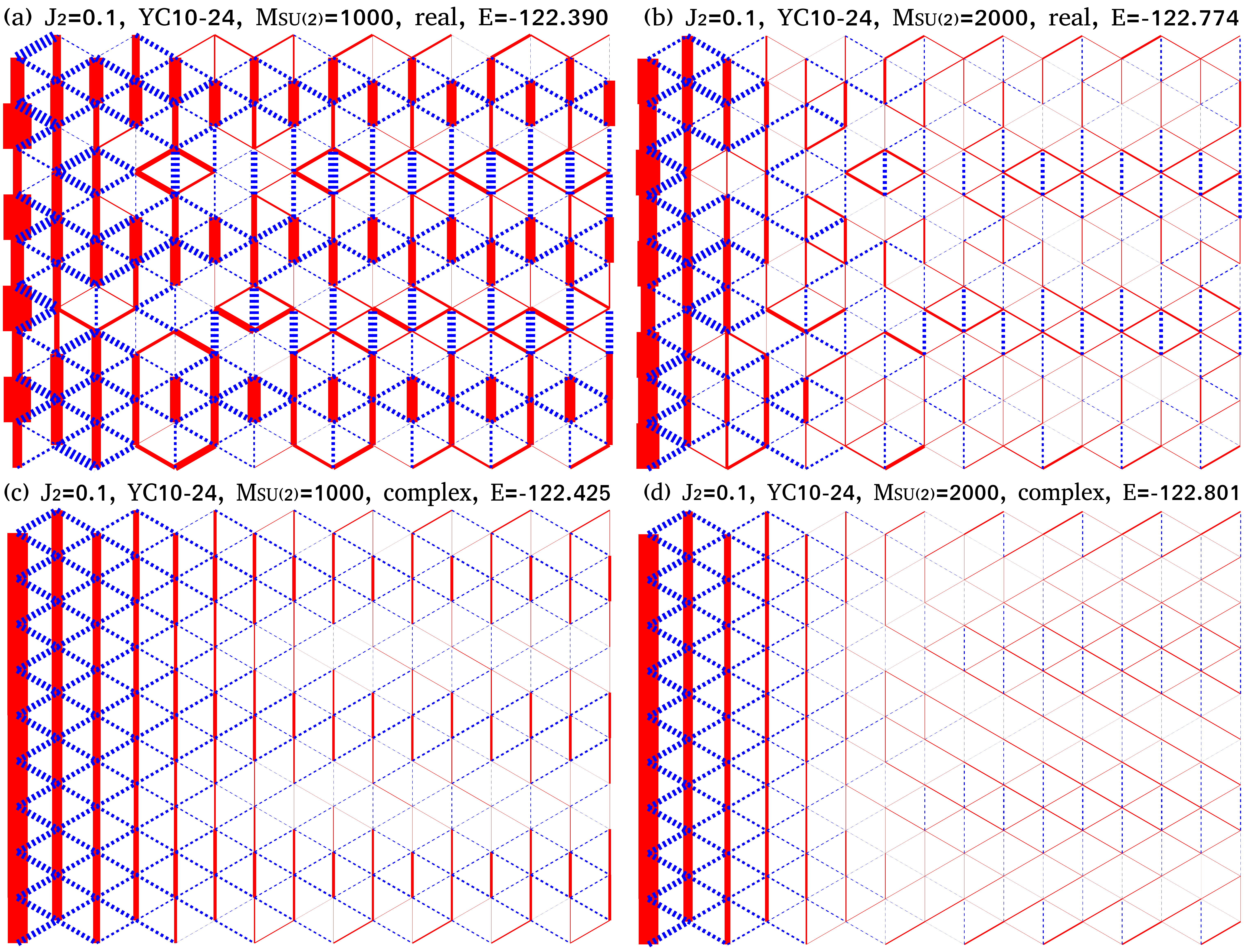}
 \caption{(color online) The NN bond textures for the even sector of $J_2 = 0.1$ on the YC10-24 cylinder 
          are calculated by using real [(a) and (b)] and complex [(c) and (d)] number DMRG. 
          Different $SU(2)$ states are kept: $M_{SU(2)}=1000$ in (a) and (c); $M_{SU(2)}=2000$ in (b) and (d).
	  The left $16$ columns are shown here. In all figures, all the bond energy have subtracted a value $-0.18$. 
	  The red solid and blue dashed bonds denote the negative and positive bond textures, respectively.
	  $E$ is the total energy for each state.}\label{suppl:bondyc10_1}
\end{figure*}

In Fig. 2 of the main text, we have shown the different NN bond energy distributions
of the ground states in different sectors on YC8 cylinder.
Here, we demonstrate the bond energy distributions for $J_2=0.1$ on YC10 cylinder
in both sectors, and compare the results in the even sector obtained from
the real and complex number states.
As the even sector is harder to converge in DMRG calculations, we first
show the bond energy distributions by keeping different states in the even sector.
As shown in Fig.~\ref{suppl:bondyc10_1}, by keeping $1000$ $SU(2)$ states,
both the real and complex states exhibit some sort of string pattern.
In particular, the real state shown in Fig.~\ref{suppl:bondyc10_1}(a) has much stronger fluctuations.
With increasing the kept states to $2000$, the fluctuations
become weaker. By comparing the ground-state energy of the real and complex
states by keeping the same optimal states such as $M_{SU(2)}=2000,3000$,
we find that the two states have the near identical bulk energy,
but the real state has the higher total energy as there are much stronger
bond energy fluctuations on the edge. This indicates that the real wave function
is less converged near the boundary.

\begin{figure}
  \includegraphics[width=0.9\linewidth]{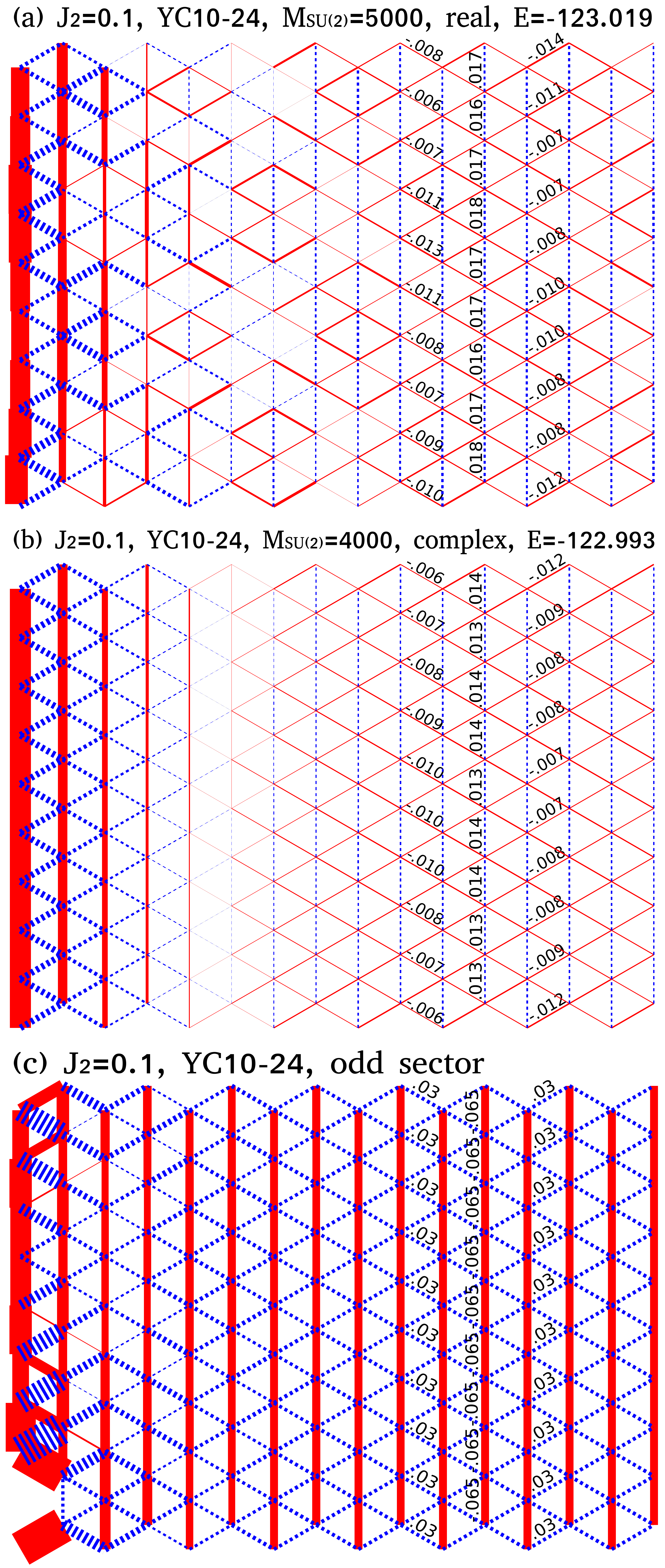}
 \caption{(color online) The NN bond textures for $J_2 = 0.1$ on YC10 cylinders. The left $16$ columns are shown here.
 	  (a) is the even sector obtained from real number DMRG by keeping $5000$ $SU(2)$ states. (b) is the even	      
	  sector obtained from complex number DMRG by keeping $3000$ $SU(2)$ states. 
	  (c) is the odd sector obtained by removing one site in
	  each boundary. In all figures, all the bond energy have subtracted a value $-0.18$. 
	  The numbers are the bond texture values. The red solid and blue dashed bonds denote the negative 
	  and positive bond textures, respectively. $E$ is the total energy. For the system (a) by keeping $4000$
	  $SU(2)$ states, the total energy $E=-122.978$, which is still higher than the energy of the complex state in (b).}\label{suppl:bondyc10_2}
\end{figure}

In Fig.~\ref{suppl:bondyc10_2},
we show the bond energy distributions obtained by keeping the most states for both the
even and odd sectors, where the fluctuations are much smaller compared with the data
in Fig.~\ref{suppl:bondyc10_1}. The real and complex states in the even sector have
the same pattern of the bond energy distribution in the bulk, which has the weaker vertical bonds.
In contrast, the vertical bonds become the stronger bonds in the odd sector.
This difference between the two sectors is consistent with our observations for YC8 cylinder, which
are shown in Fig. 2 of the main text.

\subsection{III. Nematic order}

\begin{figure}
 \includegraphics[width=\linewidth]{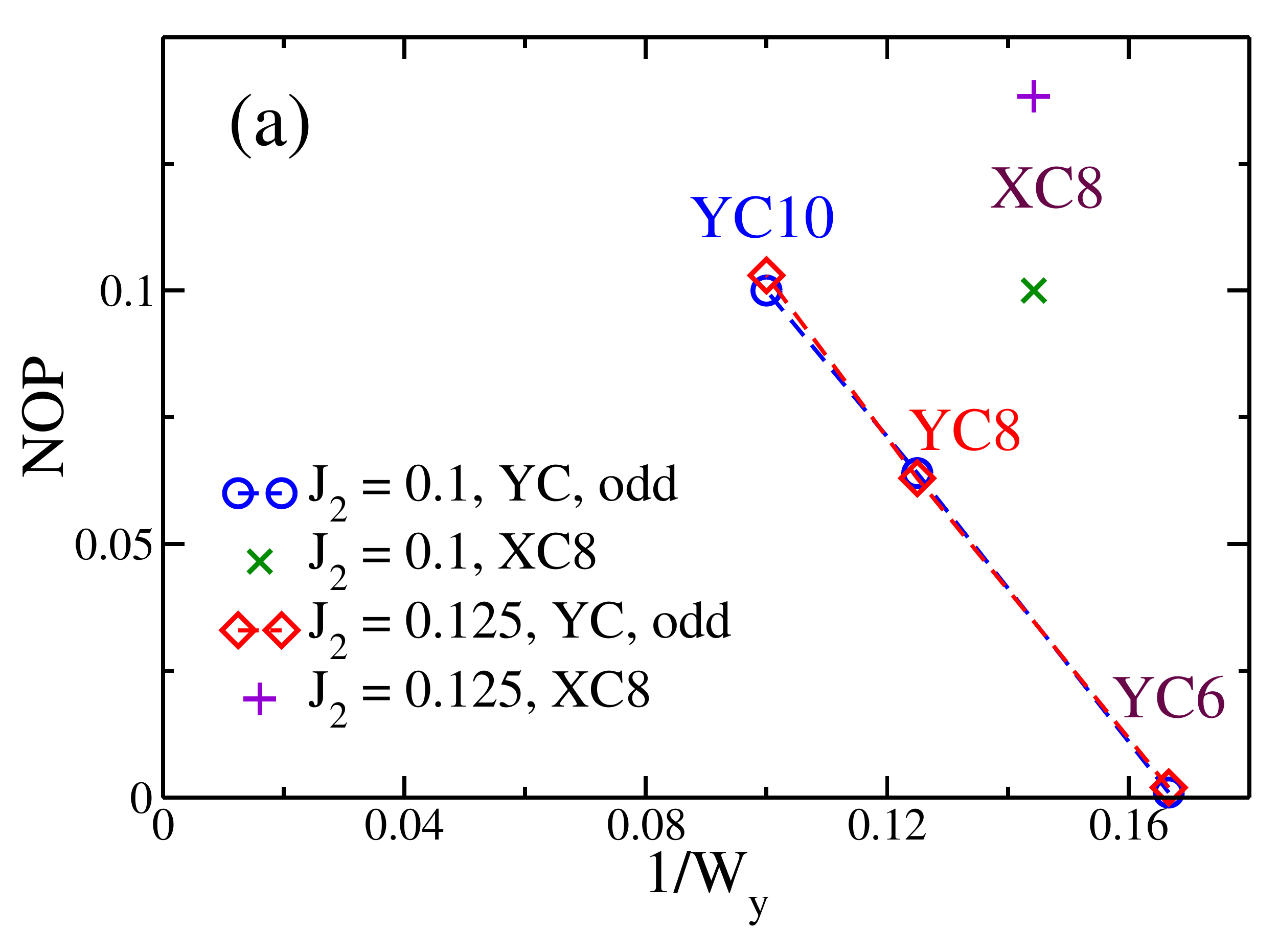}
 \includegraphics[width=\linewidth]{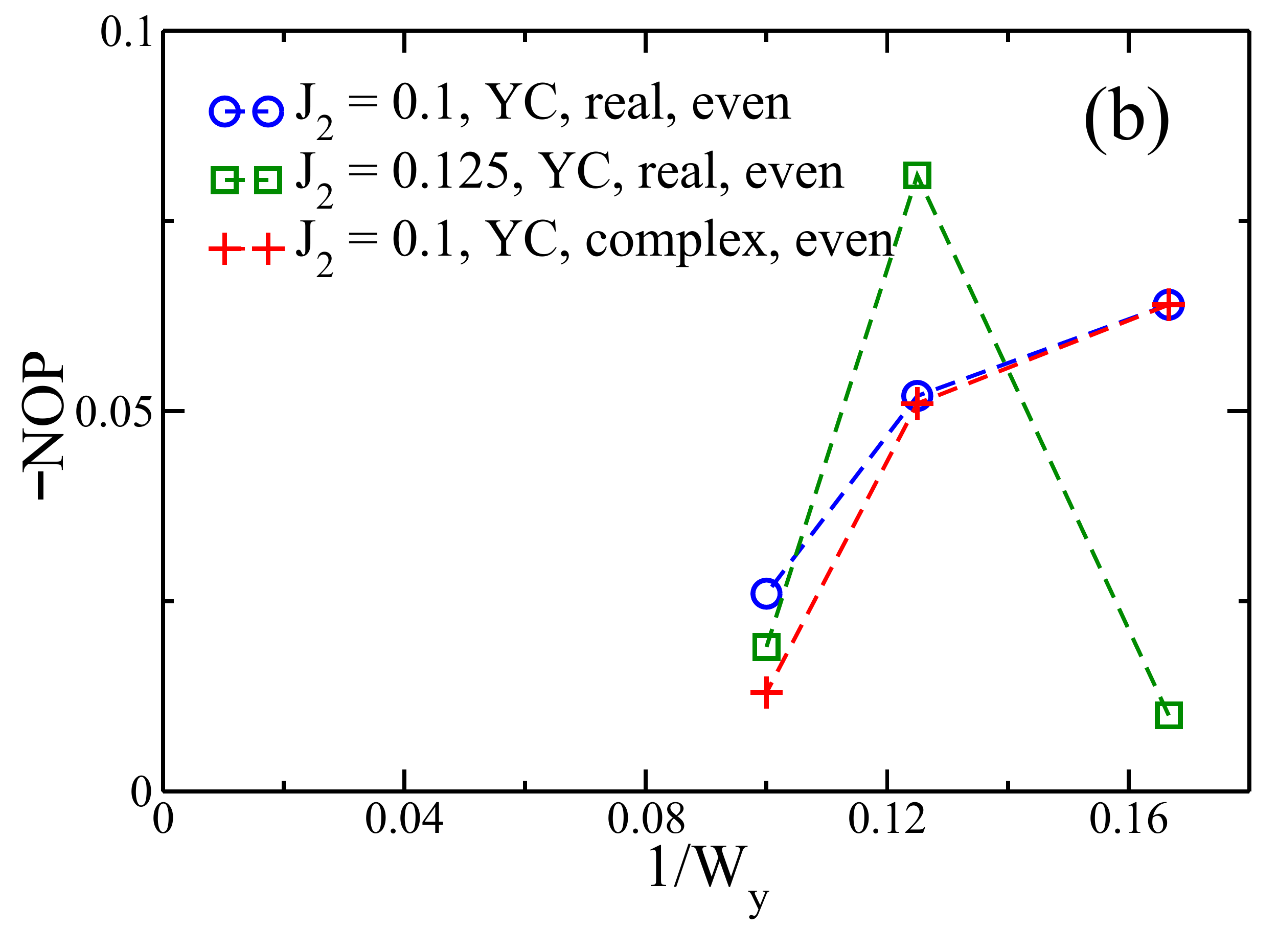}
 \includegraphics[width=\linewidth]{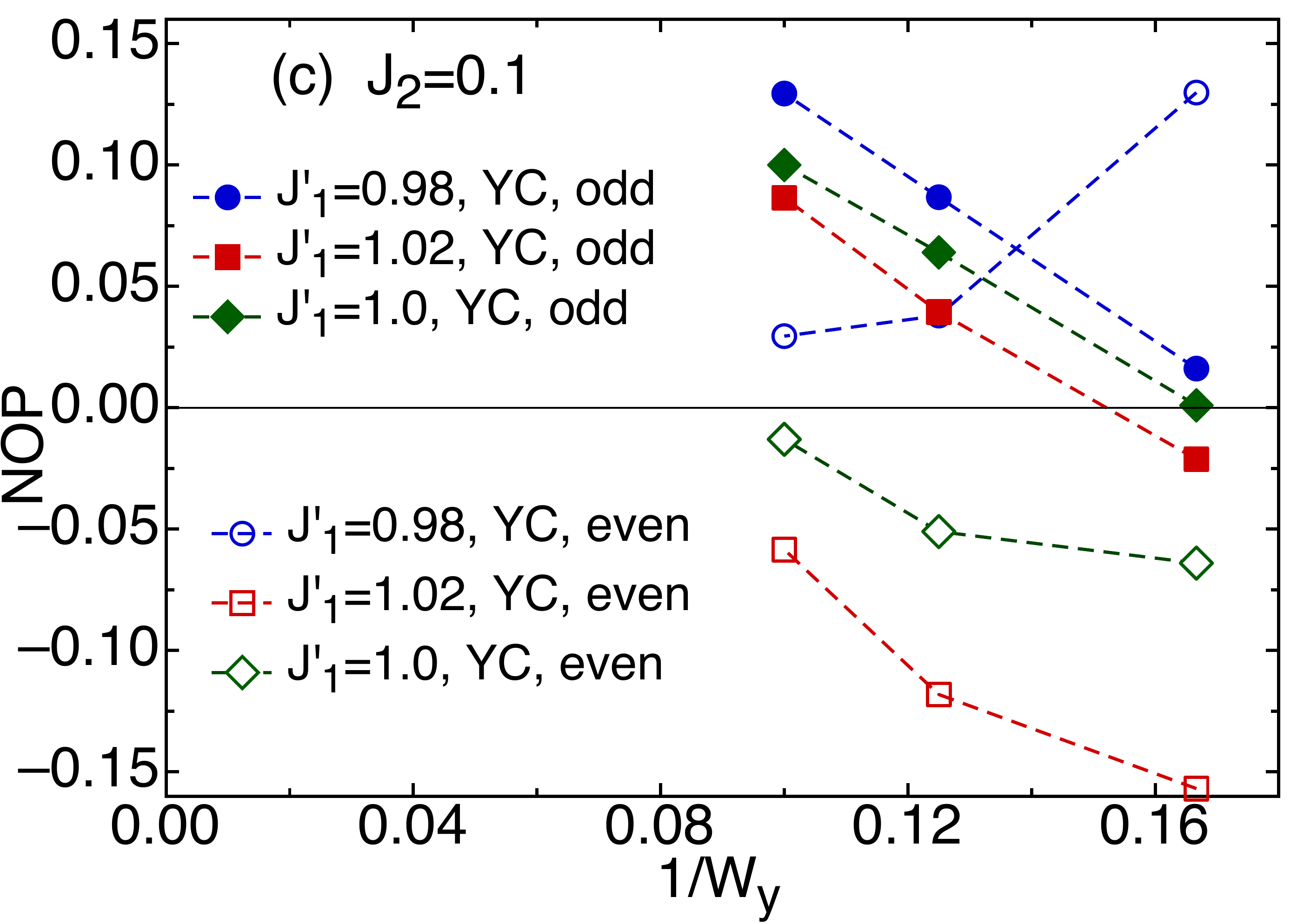}
 \caption{(color online) Cylinder width dependence of the nematic order parameter (NOP)
	  for $J_2 = 0.1$ and $0.125$ in (a) the odd sector and (b) the even sector.
	  In the odd sector (a), the NOP increases with growing system width.
	  In the even sector (b), we show the NOP obtained from both real and complex number DMRG calculations,
	  which are consistent on YC6 and YC8 cylinders. On YC10 cylinder, the complex states have the
	  better convergence and thus exhibit the smaller fluctuations.
	  (c) is the NOP in the even and odd sectors at $J_2=0.1$ 
	  with tuning the bond anisotropy $J'_{1}$.}\label{suppl:nop}
\end{figure}

In Fig.~\ref{suppl:bondyc10_2}, we notice that the NN bond energy distributions have
the lattice anisotropy with different bond energy along different lattice directions.
In particular, the bond energy in the odd sector seems to have
a strong anisotropy as shown in Fig. 2(b) of the main text and
Fig.~\ref{suppl:bondyc10_2}(c) in Supplemental Material.
The strong lattice anisotropy may indicate a lattice rotational symmetry breaking.
To investigate such a possibility, we study the anisotropy on different
cylinder systems. For convenience, we define the nematic order parameter (NOP) as the difference 
of the bond energy of the vertical bond and the others (NOP = Bond$_{\rm zigzag}$ - Bond$_{\rm vertical}$).
In Fig.~\ref{suppl:nop}, we present the NOP for $J_2=0.1$
and $0.125$ on different cylinders in two sectors, 
together with the results by tuning the anisotropic bond coupling $J'_1$.
In the odd sector, the NOP increases continuously with growing system width 
for $L_y=6,8,10$ on YC cylinder for different $J'_1$. On the XC8 cylinder, 
the systems also have strong anisotropy. 
These results may suggest a finite NOP in thermodynamic limit. 
However, in the even sector the behaviors of the NOP 
appear distinct for different $J'_1$. 
For $J'_1=1.0$ and $1.02$, the negative NOP on $L_y=6,8,10$ indicates the weaker vertical bonds.
However, for $J'_1=0.98$, the vertical bonds become the stronger bonds,
and the NOP decreases continuously with growing system width for $L_y=6,8,10$.

\subsection{IV. Entanglement Entropy}

\begin{figure}
 \includegraphics[width=\linewidth]{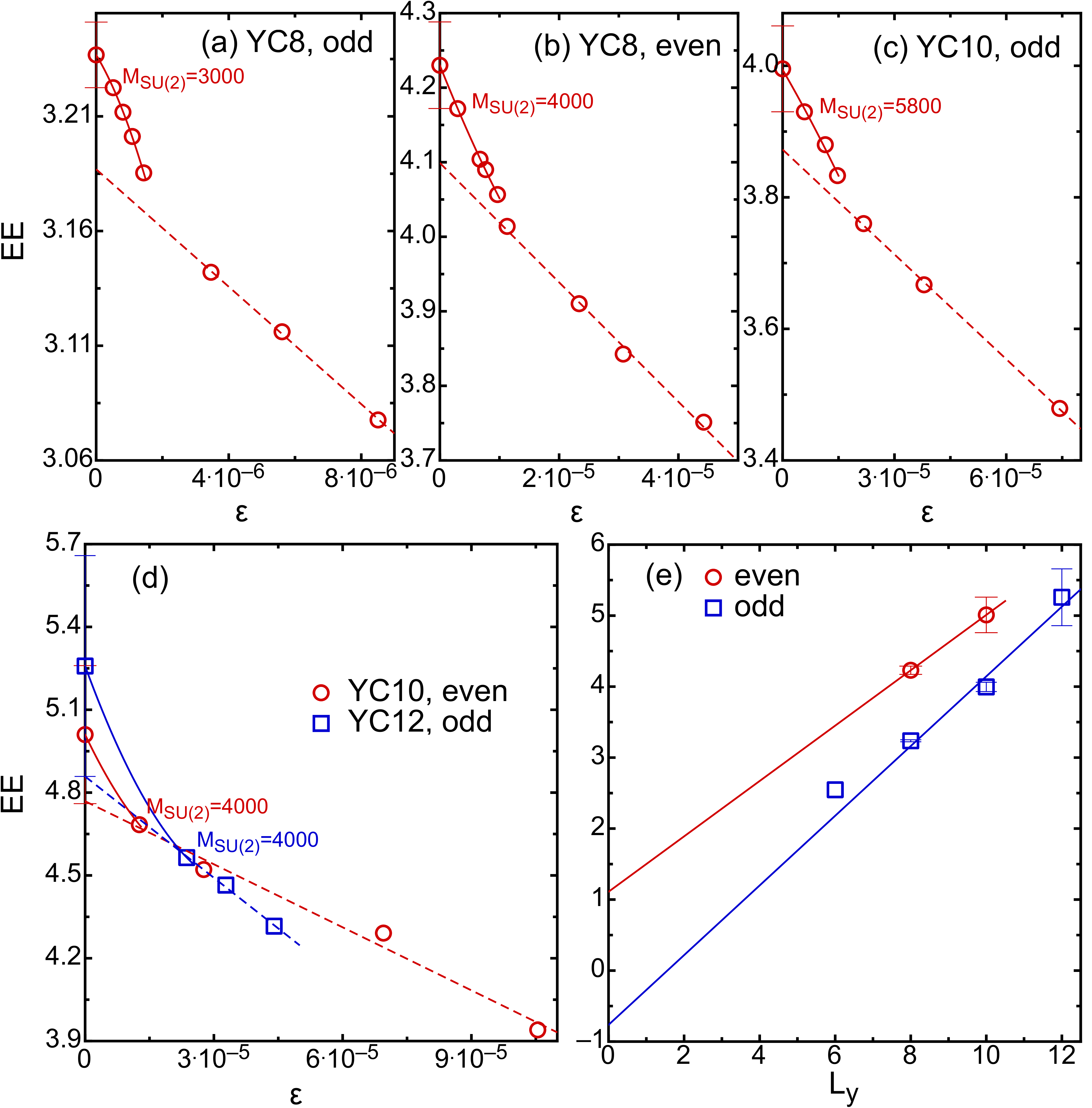}
 \caption{(color online) (a)$\sim$(d) The entanglement entropy as a function of the DMRG truncation error $\varepsilon$ at $J_2=0.1$ on YC8, YC10, and YC12 cylinders in the even and odd sectors. The largest kept $SU(2)$ DMRG states $M_{SU(2)}$ are shown. In (a)$\sim$(c), we show the convergency of the EE by increasing the kept states. By doing the linear fitting with the less accurate data (red dashed lines), the extrapolated entropy is far from convergence. With more accurate data, we do the quadratic fitting (red solid lines), and the error bar is the difference between the extrapolated EE and the last raw data. In (d) we show the EE in the even sector on YC10 cylinder and odd sector on YC12 cylinder. Even with $4000$ SU(2) DMRG states, most of these data have the truncation error larger than $2\times10^{-5}$, which indicates that they are far from convergency. Based on the experience of (a)$\sim$(c), we do the linear fitting (dashed lines) to obtain the lower bound for the EE, and also make a quadratic extrapolation (solid lines) with the guessed big error bar for these two points. (e) The width dependence of the entanglement entropy in the even and odd sectors at $J_2=0.1$.}\label{suppl:ee}
\end{figure}

We have calculated the entanglement entropy (EE) on the $L_y=6$, $8$, $10$, and $12$ YC cylinders at $J_2=0.1$. In Figs.~\ref{suppl:ee}(a)$\sim$(d), we show the entropy for $L_y=8$, $10$, and $12$ cylinders by keeping different $SU(2)$ DMRG states $M_{SU(2)}$. As shown in Fig.~\ref{suppl:ee}(a)$\sim$(c), by using the less accurate data with the big truncation error, the linear extrapolated EE on YC8 cylinder and in the odd sector on YC10 cylinder is far from the converged results. However, with the increasing kept states, the truncation error decreases, and the more accurate EE is obtained. These three figures suggest that it is hard to obtain the converged EE for the wider systems. In the even sector on YC10 cylinder and odd sector on YC12 cylinder (Fig.~\ref{suppl:ee}(d)), we have kept up to $4000$ $SU(2)$ states, however, most data have the truncation error larger than $2\times10^{-5}$, which indicates that they are far from convergency. By using these data, we only can obtain the lower bound of the EE with a linear fitting, and due to the big truncation error, it is hard to get the reliable topological entanglement entropy on large cylinders. Based on the entropy data for $L_y = 8$ and $10$ in the odd sector with the smaller error bar, we get the topological entanglement entropy close to $\ln 2$ from a reasonable fitting as shown in Fig.~\ref{suppl:ee}(e), which is consistent with the gapped $Z_2$ spin liquid.

\end{document}